\documentclass[
reprint,
amsmath, amssymb,
aps, 
]{revtex4-2}

\usepackage{graphicx}
\usepackage{dcolumn}
\usepackage{bm}
\usepackage{verbatim}
\usepackage{color}

\begin{document}

\preprint{APS/prfluids}

\title{Emergent oscillations and chaos in non-compliant microfluidic networks}

\author{Yanxuan Shao$^{1,2}$}
\author{Jean-Regis Angilella$^3$}
\author{Adilson E. Motter$^{1,2,4,5}$}
\affiliation{$^1$Department of Physics and Astronomy, Northwestern University, Evanston, IL 60208, USA\\
$^2$Center for Network Dynamics, Northwestern University, Evanston, IL 60208, USA\\
$^3$Normandie Universit\'e, UNICAEN, UNIROUEN, ABTE, Caen 14000, France\\
$^4$ Northwestern Institute on Complex Systems, Northwestern University, Evanston, IL 60208, USA\\
$^5$ Department of Engineering Sciences and Applied Mathematics, Northwestern University, Evanston, IL 60208, USA}
\date{\today}

\begin{abstract} 
Incompressible fluids in microfluidic networks with non-rigid channels can exhibit flow rate oscillations analogous to electric current oscillations in RLC circuits. This is due to the elastic deformation of channel walls that can store and release fluid, as electric capacitors can store and release electric charges. This property is quantified through the compliance of the system, defined as the volume change relative to the pressure change. In systems with rigid walls and incompressible fluid, compliance vanishes and no oscillations can occur through this mechanism. Here, we show that not only oscillations but also chaos can emerge in the flow-rate dynamics of non-compliant microfluidic networks with incompressible fluid. Notably, these dynamics emerge {\it spontaneously}, even under time-independent driving pressures. The underlying mechanism is governed by the effect of fluid inertia, which becomes relevant at moderate Reynolds numbers observed in microfluidic systems exhibiting complex flow patterns. The results are established using a combination of direct numerical simulations and a reduced model derived from modal analysis. This approach enables us to determine the onset of oscillations, the associated bifurcations, the oscillation frequencies and amplitudes, and their dependence on the driving pressures.
These findings can inspire novel studies and applications of previously unexplored oscillatory and chaotic regimes in non-compliant microfluidic systems.

\color{black}
\end{abstract}
\maketitle

\section{\label{sec:intro}Introduction}
Fostered by the need to develop versatile technologies for (bio)chemical analysis and synthesis, microfluidic systems have undergone great progress over the past decade as platforms that allow controlled manipulation of minute amounts of fluid \cite{whitesides2006origins, sackmann2014present, nguyen2019fundamentals}. These systems can offer high sensitivity and resolution within short timescales and at low costs, making them potentially applicable across various fields. 
Applications include the monitoring of cellular processes \cite{lou2024microfluidic, el2006cells}, analysis of biofilm formation \cite{karimi2015interplay},
materials fabrication and manipulation \cite{xin2019microfluidics, dong2019microfluidics},
microsynthesis \cite{liu2017microfluidics}, and
medical diagnostic testing \cite{yager2006microfluidic, pandey2018microfluidics}. To accurately control fluids in space and time, most existing microfluidic designs assume the flow to be laminar and the actuation to rely on off-chip hardware, such as externally-driven microvalves \cite{weibel2005torque, abate2008single, xia2021nonlinear} and micropumps \cite{leach2006optically, zhang2009pmma}. However, reliance on external hardware limits the affordability
and portability of microfluidics outside laboratory settings. 

To reduce dependence on external hardware, advances have been made in developing on-chip devices to control microfluidic flows. In particular, various flow patterns of interest have been generated by
controlling time-modulated pressure inputs with the aid of embedded membranes \cite{leslie2009frequency}.
Using on-chip membrane valve structures, researchers have also been able to design logic gates in pneumatic circuits \cite{grover2006development, jensen2007micropneumatic, xia2021nonlinear}, which has allowed the creation of microfluidic clocks in which a fluid flow in a microchannel oscillates under the action of pneumatic actuators \cite{duncan2013pneumatic}. This development is of substantial interest since controlled oscillators are crucial components for the development of lab-on-a-chip devices. 

Less controlled oscillations can also be induced in the absence of external modulation 
when the fluid is compressible and/or the channel walls themselves are elastic \cite{bruus2007, ahamed2025comprehensive}. 
In this case, the channels can store and release fluid much in the same way as capacitors store and release charges in electric circuits,  which gives rise to flow rate oscillations analogous to the current oscillations observed in RLC circuits. 
This mechanism has been explored to generate oscillations in complex microfluidic networks \cite{ruiz2020topologically, ruiz2021emergent}. 
The effect can be quantified in terms of hydraulic capacitance, or compliance, defined as the derivative of the volume with respect to the pressure \cite{bruus2007, christov2021soft}.
While even simple systems can be designed to be compliant, fabrication of devices with elastic walls is generally involved. Moreover, although energy can also be stored through fluid compression, many applications require the use of an incompressible working fluid.

A longstanding question has been whether oscillations are possible for incompressible fluids in systems that are both unmodulated {\it and} non-compliant. Such behavior is generally regarded as unexpected since it is not immediately clear how the system could store and release energy to generate sustained oscillations. 
Yet, it has been recently shown that  microfluidic networks with rigid microchannels and incompressible fluids are capable of exhibiting spontaneous flow rate oscillations under the sole effect of a constant driving pressure \cite{case2020spontaneous}. 
This initial prediction was for designs with strongly nonlinear obstacle-laden channels exhibiting the quadratic Forchheimer flow-pressure relationship commonly observed in porous media \cite{anbari2018microfluidic}.
These flow rate oscillations are markedly distinct from oscillations in fluid trajectories arising from the Coand\u{a} effect, which are explored, for example, in oscillating feedback micromixers \cite{wang2024short, xie2017numerical}.
Crucially, the oscillations considered in this study pertain specifically to flow rates and emerge from a different physical mechanism.

In this work, we show that spontaneous microfluidic oscillations can be observed in simpler designs and described by a reduced dynamical model with only two degrees of freedom. 
The key ingredient for this behavior is weak hydrodynamic nonlinearity created by inertial effects at moderate Reynolds numbers. Physically, the oscillations are associated with the formation and evolution of vortices.
Our model allows us to characterize the onset of oscillations explicitly in terms of Hopf bifurcations.  Using this model in combination with rigorous direct numerical simulations (DNS) of the Navier-Stokes equations, we identify and characterize a wide range of driving pressures for which spontaneous periodic oscillations occur. Through the inertial-governed mechanism, we demonstrate that spontaneous oscillations are possible even in microfluidic networks with a rather simple geometry, which is important in connection with applications. 
Serialization of these networks shows that oscillations can persist and remain synchronized in all channels while also becoming increasingly complex and eventually chaotic as a function of the driving pressures. 

This previously unappreciated mechanism for generating oscillations can advance ongoing efforts to create built-in microfluidic clocks of easy fabrication and without dependence on external hardware. 
A related nonlinear effect has been shown to generate negative resistance (i.e., a decrease in flow rate upon an increase in pressure difference) \cite{case2020spontaneous, martinez2024fluidic}, which is another property of relevance for on-chip control that was previously observed in elastic, externally actuated  networks \cite{martinez2024fluidic}.
More broadly, this research adds to a growing literature on the role of   nonlinearities induced by geometry and barriers in microfluidic networks \cite{case2019braess, battat2022nonlinear}.

The paper is organized as follows. Section \ref{sec:network} introduces the geometry and dynamics of an elementary microfluidic network designed to exhibit oscillations, along with a DNS characterization of the timescales and topology of the flow.  
Section \ref{sec:model} presents a modal analysis of the DNS data, which is used to create the reduced model.
Both the simulations and model predict the emergence of periodic oscillations, which allows us to create comprehensive bifurcation diagrams. Section \ref{compo_in_series} addresses the behavior of such elements coupled in series, showing that serialization can give rise to complex periodic and aperiodic dynamics due to the excitation of higher order modes. 
Discussion and conclusions are presented in Section~\ref{conclusion}. 

Throughout, we assume the fluid channels to be two-dimensional and the working fluid to model water.  Two-dimensional fluid models approximate fluid behavior by assuming an infinite third dimension and neglecting variations along it. These models provide valuable insights while significantly reducing the computational cost compared to three-dimensional simulations, which has contributed to their widespread use in the field.

\begin{figure}[h]
\centering
\includegraphics[width=0.4\textwidth]{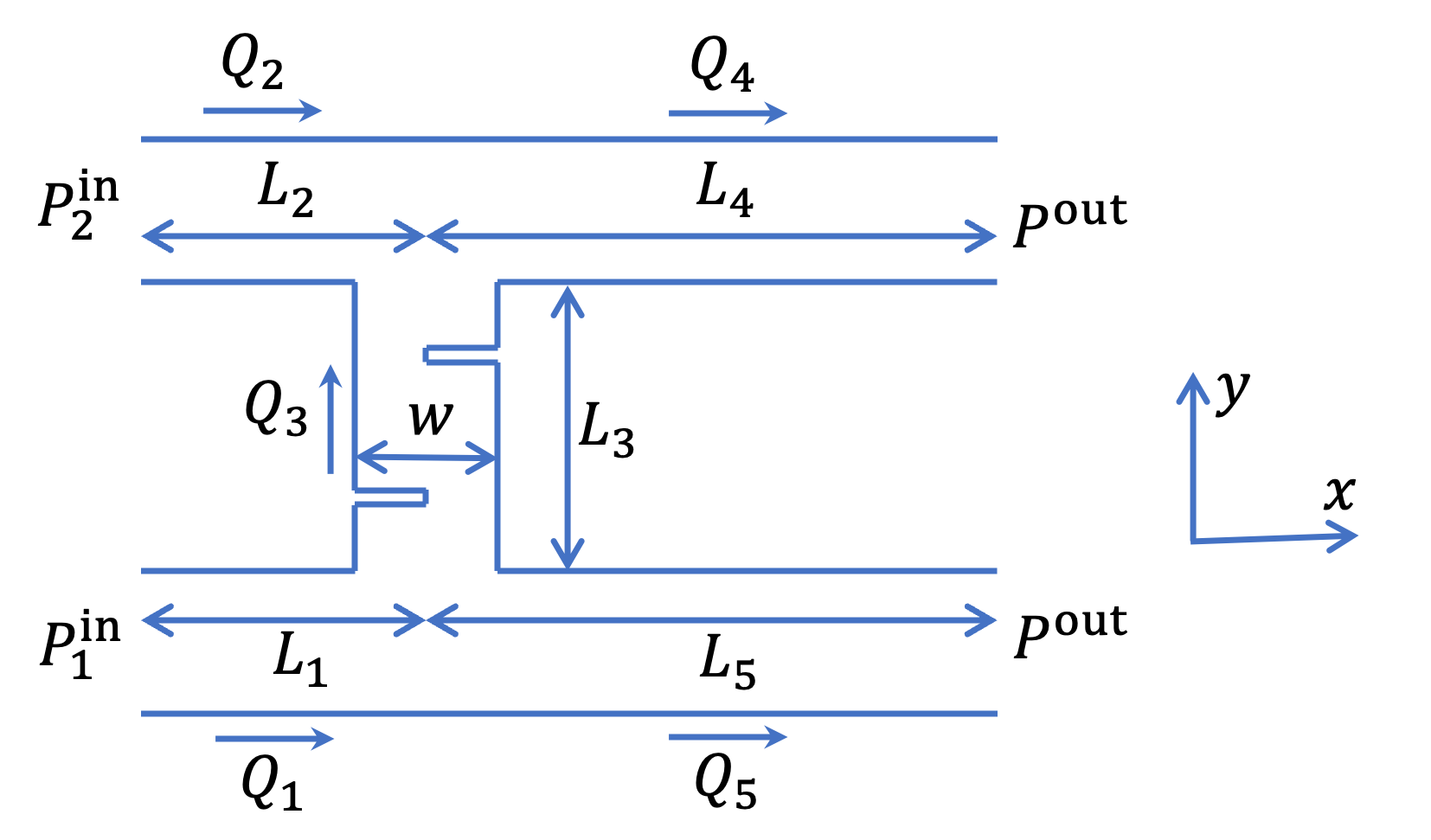}
\caption{ 
Geometry of the elementary microfluidic network, where the width of all the channels is $w=500\, \mu \text{m}$ and the lengths are $L_1=L_2=L_3=1\, \text{mm}$ and $L_4=L_5=2\, \text{mm}$. The length and width of the blades in the connecting channel are $w/2$ and $w/10$, respectively. The two blades are symmetrically positioned and at a vertical distance $w$  from each other. 
The inlet pressures $P_1^{\text{in}}$ and $P_2^{\text{in}}$ can be tuned independently and are assumed to be larger than the outlet pressure $P^{\text{out}}$. 
}
\label{network}
\end{figure}

\section{\label{sec:network}   Numerical simulations and the emergence of oscillations}

We first study the dynamics of the fluid flow for the two-dimensional microfluidic network shown in Fig.\ \ref{network}.
The network is composed of two straight channels connected by a   transversal channel with two
blade-shaped barriers.  This results in a network with five channel segments of generally different lengths $L_j, \, j=1,...,5$, but all assumed to have the same width $w$.
The network has two inlets, two junctions, and two outlets.
The blades in the connecting channel induce nonlinearity, as discussed below.
This design further simplifies the system in Ref.\ \cite{case2020spontaneous}, where obstacle-loaded channels were used as a main source of nonlinearity.
For the pressures considered in our analysis, the fluid in the longitudinal channels moves from the left to the right, and we adopt the convention that the flow rates, denoted $Q_j$, are positive for the directions indicated by the arrows in Fig.\ \ref{network}.
For the entire network, we only control the static pressures at the two inlets, $P_1^{\text{in}}$ and $P_2^{\text{in}}$, which are constrained to be time independent; the outlets are assumed to experience a common constant pressure $P^\text{out}$, which is set to zero.
This boundary condition is consistent with pressure-driven flow microfluidics in applications where the system inlets are connected to pressurized fluid reservoirs and the outlets are at (static) atmosphere pressure.

\begin{figure*}[th]
\centering
\includegraphics[width=0.9\textwidth]{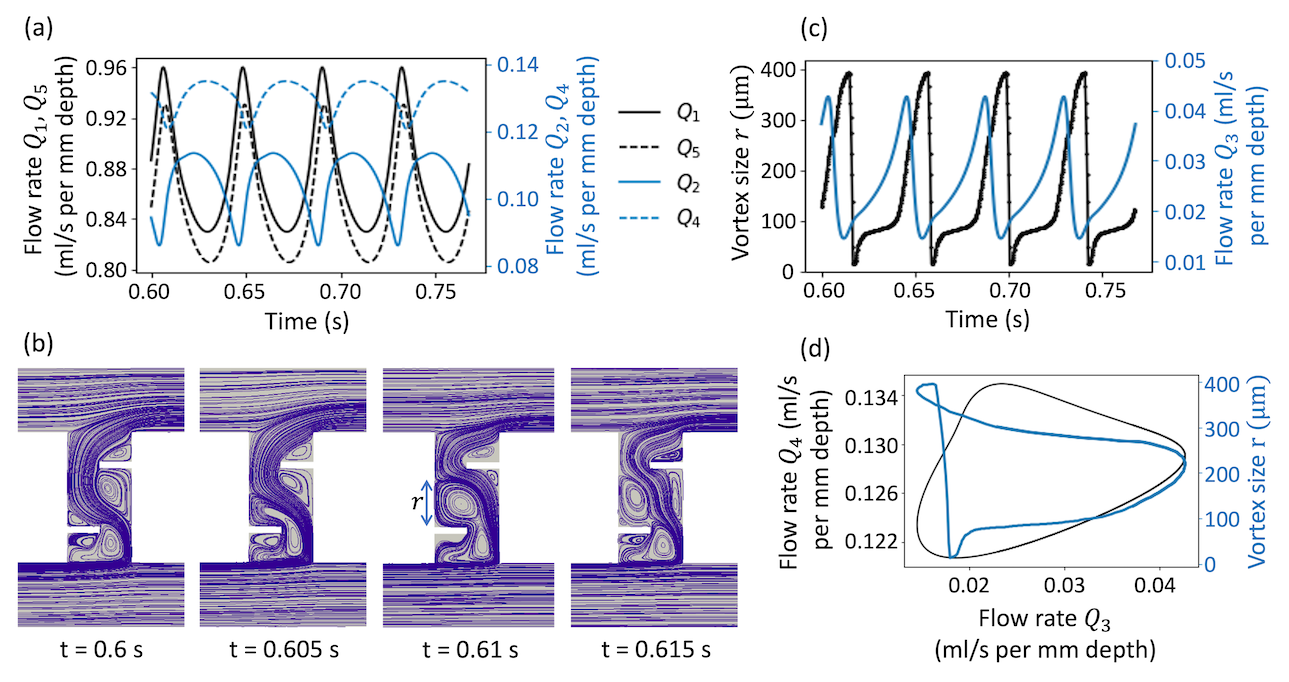}
\caption{
Representative simulation showing spontaneous oscillations 
in the network introduced in  Fig.\ \ref{network}
when the inlet pressures are $P_1^{\text{in}}=80\, \text{Pa}$ and $P_2^{\text{in}}=60\, \text{Pa}$. 
(a) Time-periodic oscillations of the flow rates $Q_1$ and $Q_5$ (left axis) and $Q_2$ and $Q_4$ (right axis). (b) Snapshots of the streamlines within and around the connecting channel visualized using ParaView at four different times of an oscillation period. The streamlines are marked by first choosing random points in the domain and then tracing the contours of the velocity field passing through those points at the given time. (c) Oscillations of the vortex size $r$ (left axis) and flow rate $Q_3$ (right axis). (d) Limit cycles of the oscillations in the phase spaces of the flow rate $Q_4$ (left axis) and vortex size $r$ (right axis) vs. flow rate $Q_3$.
}

\label{osci_P1eq80-P2eq60}
\end{figure*}

To determine the fluid flow in this elementary microfluidic network, we perform {DNS} of the incompressible Navier-Stokes equations in two dimensions,
\begin{eqnarray}
\label{NavierStokesEq}
    \rho \frac{\partial \textbf{u}}{\partial t} + \rho(\textbf{u}\cdot \nabla)\textbf{u}&=& -\nabla p + \mu \Delta \textbf u   , \\
    \nabla \cdot \textbf u &=& 0,
    \label{Incompress}
\end{eqnarray}
where $\textbf u$ denotes the velocity field, $p$ is the pressure field, $\rho$ is the density of the fluid, and $\mu$ is the dynamic viscosity. The Reynolds number---which can be interpreted as the ratio of the order-of-magnitude of the inertial term $\rho(\textbf{u}\cdot \nabla)\textbf{u}$ to the viscous term $ \mu \Delta \textbf u $---reads Re $=\rho U_0 w/\mu$, where $U_0$ is a typical value of the flow speed. In the following, $U_0$ will be taken to be the cross-sectional average velocity. 

The numerical solutions of Eqs.\ (\ref{NavierStokesEq}) and (\ref{Incompress}) are 
obtained using OpenFOAM (version 5.0). Specifically, we used the PisoFoam solver for time-dependent simulations and the SimpleFoam solver for steady-state simulations. We consider the fluid to be {water-like} with density $\rho=10^3\: \text{kg}/\text{m}^3$ and dynamic viscosity $\mu = 10^{-3}\, \text{Pa}\!\cdot \!\text{s}$. The fluid is regarded as incompressible and no-slip boundary conditions are assumed at the walls. The meshes used in the simulations were created using Gmesh with minimum and maximum cell sizes of $9\: \mu \text{m}^2$ and  $64\: \mu \text{m}^2$, respectively. 
 For the simulations considered in this study, 
the typical Reynolds numbers  exhibiting flow rate oscillations are around one hundred for the top channels, a few tens for the connecting channel (ranging from 27 to 40), and several hundred for the bottom channels.  

The DNS show that periodic flow rate oscillations {\it emerge spontaneously} for a range of parameters in this system, even though the inlet (and outlet) pressures are fixed. 
Figure \ref{osci_P1eq80-P2eq60} 
presents an example of the dynamics when the inlet pressures are $P_1^{\text{in}}=80\,\text{Pa}$ and $P_2^{\text{in}}=60\,\text{Pa}$.
The oscillatory dynamics of the inlet and outlet flows are shown in Fig.~\ref{osci_P1eq80-P2eq60}(a). 
The oscillations are induced by recirculation cells (vortices) in the connecting channel, which result from the deformation of the streamlines caused by the barriers. Crucially, the shape and size of the vortices vary in time
[Fig.\ \ref{osci_P1eq80-P2eq60}(b)].  
To quantify the relation between vortex dynamics and flow rate, we choose to measure the distance $r(t)$ between the stagnation point on the left boundary of the connecting channel and the upper wall of the left blade [as indicated in Fig. \ref{osci_P1eq80-P2eq60}(c)]. For simplicity, $r(t)$ will be referred to as ``vortex size'' throughout the paper.
The dynamics of the vortex size $r(t)$ are summarized in  Fig.\ \ref{osci_P1eq80-P2eq60}(c), showing
that it varies periodically and with the same period for the parameters under consideration. This indicates that the underlying solutions converge to limit cycles, as shown explicitly in Fig.\ \ref{osci_P1eq80-P2eq60}(d).

It follows from Figs.\ \ref{osci_P1eq80-P2eq60}(c)-(d) that the flow rate $Q_3$ peaks when the vortex is at an intermediate size. At such time points, the left and right side vortices have comparable sizes and balance each other, allowing the flow to circumvent the barriers in the connecting channel without experiencing significant curvature or constriction. In contrast, when the size of the left vortex peaks, it becomes comparable to the width of the channel, and the streamlines passing through the channel are constrained to the right corner, which limits the flow rate and causes $Q_3$ to reach its minimum. Note that the role played by the two vortices is not symmetric, since the flow through the connecting channels (albeit oscillatory) is always upwards for the given inlet pressures. 

\begin{figure*}[th]
\centering
\includegraphics[width=0.8\textwidth]{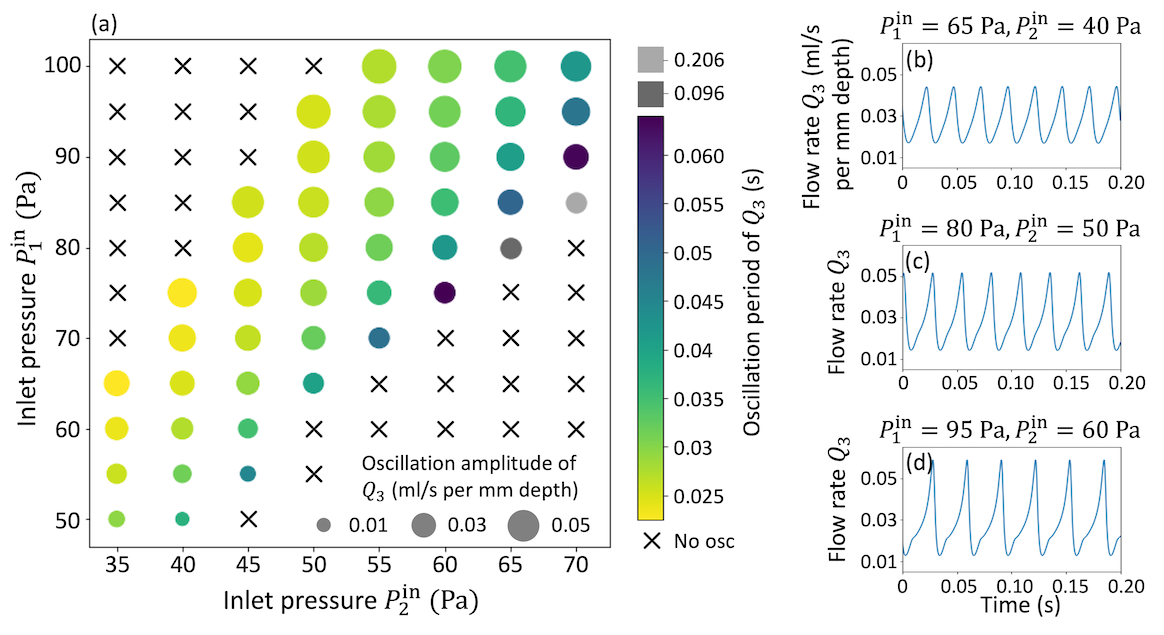}
\caption{Simulation results for the microfluidic network in Fig.\ \ref{network}. (a) Diagram of the periods and amplitudes of the oscillations in the flow rate $Q_3$ as functions of the inlet pressures $P_1^{\text{in}}$ and $P_2^{\text{in}}$. The circles mark the oscillatory region, with the color and size representing the periods and amplitudes of the oscillations, respectively. The upper left and lower right regions with cross signs correspond to regimes in which no oscillations occur. (b-d) Example time series of the oscillations in the flow rate $Q_3$ for three different choices of inlet pressures. }
\label{cfd_singlechip}
\end{figure*}

To systematically examine how the periods and amplitudes are associated with the inlet pressures, Fig.\ \ref{cfd_singlechip} shows the results of direct simulations in the $(P_1^{\text{in}},  P_2^{\text{in}})$ parameter space. Spontaneous oscillations emerge for a range of pressures in the region $P_1^{\text{in}} > P_2^{\text{in}}$, where the asymmetry in the roles played by $P_1^{\text{in}}$ and $P_2^{\text{in}}$ is due to the asymmetry of the blades.
As shown in Fig.\ \ref{cfd_singlechip}(a), where the circles indicate the  properties of the flow rate oscillations in the connecting channel, the oscillations only vanish when
$P_1^{\text{in}}$ is substantially larger than $P_2^{\text{in}}$. The latter occurs because $Q_3$ is then stabilized by the high pressure drop along the connecting channel. 
The regions with no oscillations are indicated by the crosses.
When $P_2^{\text{in}}$ is fixed and $P_1^{\text{in}}$ is increased, oscillations first appear with low frequencies and small amplitudes and then grow stronger and more dynamic (larger amplitudes and generally higher frequencies). Over the  range of $P_2^{\text{in}}$ considered, the upper boundary of the oscillatory region is then marked by an  abrupt transition from high-frequency, high-amplitude oscillations to non-oscillatory behavior as $P_1^{\text{in}}$ is further increased.
The highest frequencies appear for smaller $P_2^{\text{in}}$ while higher amplitudes can occur for larger  $P_2^{\text{in}}$.  
The range of $P_1^{\text{in}}$ for which oscillations emerge increases for larger $P_2^{\text{in}}$.
The time series of $Q_3$ for a selection of inlet pressures are shown in Fig.~\ref{cfd_singlechip}(b)-(d).
 
\vspace{0.5cm}
\section{Reduced model and analysis of Hopf bifurcations}
\label{sec:model} 
\subsection{Construction of the reduced model}

To characterize the flow structure and calculate the periods and amplitudes of the oscillations, we derived a reduced model based on modal analysis. The fluid velocity at the position $\textbf x$ and time $t$ is expanded in the form
 \begin{equation}     
\mathbf u(\mathbf x, t) =  \overline{\mathbf u}(\mathbf x) + \sum_{n=1}^\infty a_n(t) \mathbf\Phi_n(\mathbf{x}), 
 \label{PODdecomp}
 \end{equation} 
 where the bar denotes the time average. The coefficients of the decomposition satisfy $\overline{a_n(t)} = 0$. The basis functions $\mathbf \Phi_n$ are obtained using a proper orthogonal decomposition (POD) of the simulated data presented in the previous section. Similar decompositions have been widely used for turbulent flows \citep{lumley1967,berkooz1993} and have received attention also in the context of microfluidics \citep{wang2017,yang2023}. Modes are defined as the vector fields $\mathbf \Phi$ that maximize 
$ \overline{|\langle\mathbf u-\overline{\mathbf u},\mathbf\Phi\rangle|^2 } $
with $ \langle\mathbf\Phi , \mathbf\Phi\rangle = 1$, where 
$|\cdot|$ denotes absolute value and
$\langle\cdot,\cdot\rangle$ is the inner product in the space of square-integrable functions. One can check that these modes are eigenvectors of 
a self-adjoint compact operator and form a basis of orthonormal functions $\mathbf \Phi_n$ with associated eigenvalues
$\lambda_n = \overline{|a_n(t)|^2 }$, 
$n = 1,2, ...$.
Therefore, the eigenvalue $\lambda_n$ corresponds to the energy of mode $n$. The eigenvalues and eigenvectors were computed using the snapshot method \cite{sirovich1987} for $100$--$215$ snapshots of the vector field and a spatial resolution of up to $21\, 600$ grid points.

This modal decomposition provides two important pieces of information. First, it allows us to calculate the number of effective degrees of freedom of the system, obtained here as the number of POD modes with an energy $\lambda_n$ above a given threshold. The spatial patterns of the most energetic modes will then provide information about the so-called ``coherent structures'' that play a significant role in the dynamics. Second, the POD method provides a reduced model that will allow us to calculate the modal coefficients $a_n(t)$ for all times. As shown below, this is achieved by solving a dimension-reduced system of coupled ordinary differential equations with a limited number of degrees of freedom.

\begin{figure*}[th]
    \centering
    \includegraphics[width=0.95\textwidth]{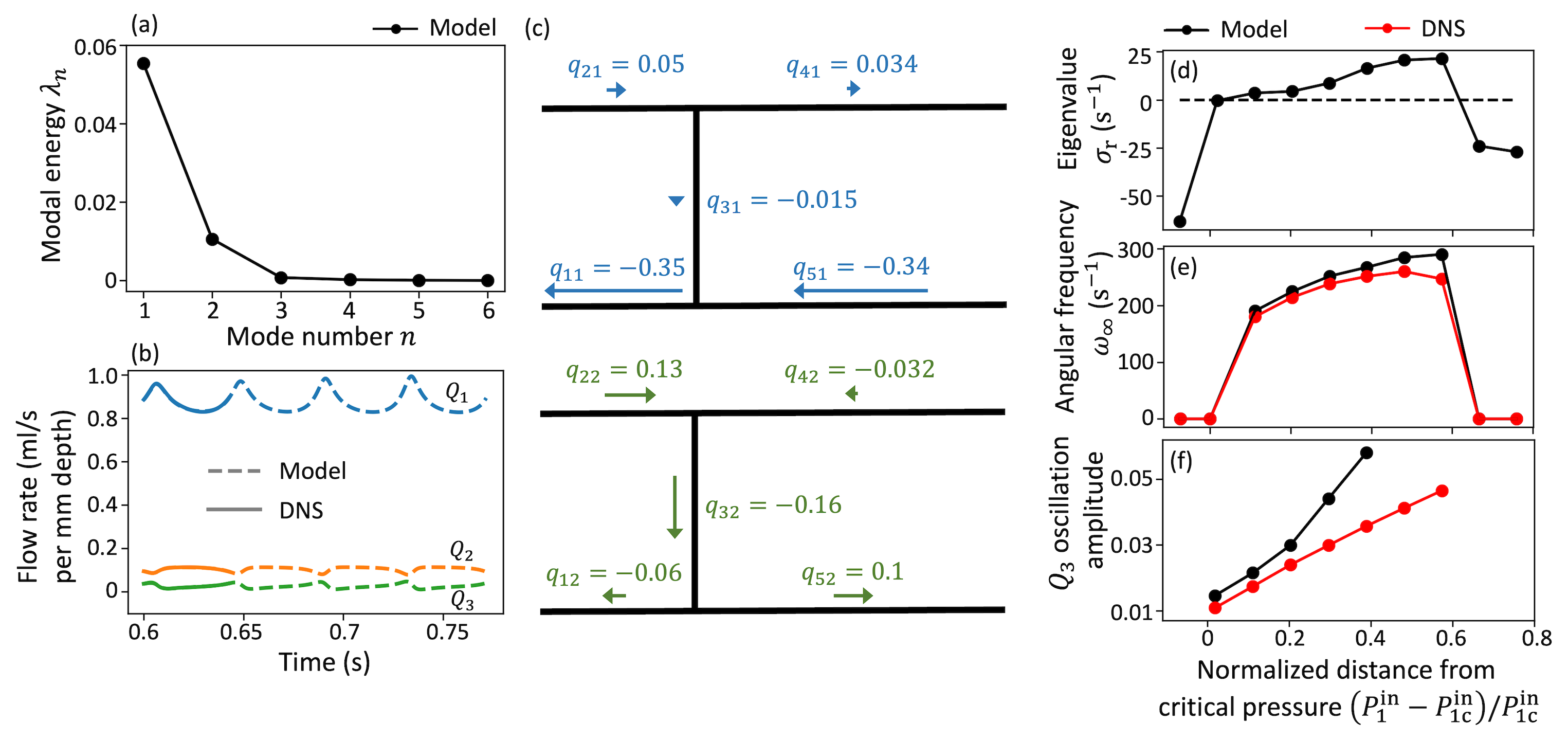}
    \caption{ Analysis of the reduced model for the elementary network. 
    (a) Modal energy (non-dimensional) for $P_1^{\text{in}}=80$ Pa and $P_2^{\text{in}}=60$ Pa. (b) Flow rates obtained by DNS (solid lines) and the reduced model (dashed lines). (c) Elementary flow rates mode 1 (top) and mode 2 (bottom) for
    $P_1^{\text{in}}=80$ Pa and $P_2^{\text{in}}=60$ Pa. 
    (d-f) Instability in the steady state solution and consequent emergence of oscillations in the reduced model
    as $P_1^{\text{in}}$ is increased beyond $P_{1c}^{\text{in}}\approx 54 $ Pa   for $P_2^{\text{in}}=45$ Pa. This transition is characterized in terms of the real part of the eigenvalue $\sigma_{\text{r}}$ of the equilibrium point (d), angular frequency $\omega_\infty$ of the saturated mode in Eq.\ (\ref{winf}) (e), and
    amplitude of the $Q_3(t)$ oscillations obtained from the solution of the normal form of the model in Eq.\ (\ref{zetainf}) (f).
    In panels (e-f), the DNS results are presented in red.}
    \label{fig:model}
\end{figure*}

Once the modal coefficients $a_n(t)$ are known, the flow rates per unit depth in any channel are obtained by integrating Eq.\ (\ref{PODdecomp}) over the channel's cross-section:
\begin{equation}
Q_j(t) = \overline{ Q_j} + \sum_{n} q_{jn} \, a_n(t),
\label{Qiai}
\end{equation}
where $q_{3n} = \int_{\Gamma_3} \mathbf{\hat{e}}_y \cdot \mathbf \Phi_n  (x,y_3)  dx$ for the connecting channel, $q_{jn} = \int_{\Gamma_j} \hat{\mathbf{e}}_x\cdot \mathbf\Phi_n(x_j,y)dy$, and
 $\hat{\mathbf{e}}_x$ and $\hat{\mathbf{e}}_y$ are the unit vectors in the respective directions. 
Since modes correspond to velocity fields, each factor $q_{jn}$ can be interpreted as the flow rate of mode $n$ through the cross-section $\Gamma_j$.

The reduced model is obtained by inserting the decomposition (\ref{PODdecomp}) into the weak form of the non-dimensional Navier-Stokes equations, leading to:
\begin{align}
& \dot a_n(t) + \underbrace{E_n  
+  \sum_s C_{ns} a_s}_{\mbox{mean-flow forcing}}
 ~~ + \underbrace{\sum_{\ell,s} A_{n\ell s} a_\ell a_s}_{\mbox{inter-mode interactions}} \nonumber \\
 & = \underbrace{P_1^{\text{in}} \, q_{1n} + P_2^{\text{in}} \, q_{2n}}_{\mbox{inlet-pressures forcing}} + \underbrace{\frac{1}{\text{Re}} D_n
 + \frac{1}{\text{Re}}  \sum_q B_{ns}  a_s}_{\mbox{viscous effects}}, 
 \label{eq:reduced_model}
 \end{align}
where
$A_{n\ell s}=\langle  (\nabla \mathbf\Phi_s )   \mathbf \Phi_\ell   , \mathbf\Phi_n \rangle$, 
$B_{ns} =   \langle\Delta\mathbf  \Phi_s , \mathbf\Phi_n \rangle$, 
$C_{ns} = \langle  (\nabla \overline{\mathbf u})   \mathbf\Phi_s  , \mathbf\Phi_n \rangle + \langle (\nabla \mathbf\Phi_s )   \overline{\mathbf u}  , \mathbf\Phi_n \rangle$, 
$D_n = \langle\Delta\overline{\mathbf u} , \mathbf\Phi_n \rangle$, 
and $E_n = \langle(\nabla \overline{\mathbf u}) \overline{\mathbf u}  , \mathbf\Phi_n  \rangle$. 
The notation $\nabla \mathbf{f}$ indicates the Jacobian matrix of $\mathbf{f}$ and  $\Delta=\nabla \cdot \nabla$ is the Laplacian operator.
Using Green's formula, $\mathbf\Phi_n=\mathbf 0$ at the walls, and $\partial \mathbf\Phi_n /\partial x = 0$ at the inlets and outlets, one obtains a first-order form for the coefficients arising in the viscous terms:
$B_{ns}  = - \langle\nabla \mathbf  \Phi_s, \nabla\mathbf  \Phi_n\rangle$ and $D_n =  - \langle\nabla \overline{\mathbf u}, \nabla\mathbf  \Phi_n\rangle$, where the inner product is now between matrix functions. These coefficients can be calculated by substituting realizations of $\mathbf u$ into these expressions. They can also be obtained using a calibration method \cite{cordier2010}, which is the method used in this paper.
In the derivation of Eq. (\ref{eq:reduced_model}), the Navier-Stokes equations are set non-dimensional by means of the density $\rho$, the length scale $w$, and an arbitrary velocity scale $u_0$. 
The Reynolds number remains as defined in section \ref{sec:network}.

This reduced model can be used to characterize the instability leading to the appearance of the periodic dynamics. When two degrees of freedom are used, the model is of the form:
\begin{align}
\label{dota1}
\dot a_1(t) = &-A_{111} a_1^2 - A'_{121} a_1 a_2 - A_{122} a_2^2 \nonumber \\
&- C'_{11} a_1 - C'_{12} a_2 + H_1  ,\\
\dot a_2(t) = & -A_{211} a_1^2 - A'_{221} a_1 a_2 - A_{222} a_2^2 \nonumber\\
&- C'_{21} a_1 - C'_{22} a_2 + H_2,
\label{dota2}
\end{align}
where $A'_{n21}=A_{n21}+A_{n12}$,   $C'_{ns}=C_{ns}+B_{ns}/$Re, and $H_n=P_1^{\text{in}} \, q_{1n} + P_2^{\text{in}} \, q_{2n} - E_n$.  This system can be re-written in vector form as
\begin{equation}
\dot {\mathbf X} = \mathbf N(\mathbf X ,\mathbf X) + \mathbf L \mathbf (\mathbf X) + \mathbf H,
\label{dotX}
\end{equation}
where $\mathbf X=(a_1,a_2)^T$,    $\mathbf H = (H_1,H_2)^T$, $\mathbf L$ is a linear function, and $\mathbf N$ is a bilinear function corresponding to the inertial term in Eq.~(\ref{NavierStokesEq}).

Figure \ref{fig:model}(a) shows the eigenvalues obtained by means of the snapshot method in the case of the chip with a single component shown in Fig.\ \ref{osci_P1eq80-P2eq60} ($P_1^{\text{in}}=80$ Pa and $P_2^{\text{in}}=60$ Pa). 
Lengths are scaled by the channel width, $w = 500\, \mu$m, and velocities are scaled by $u_0=1$ m/s, which is of the order of magnitude of the characteristic velocity observed in our simulations.  The eigenvalue $\lambda_n$ 
decreases rapidly with $n$, indicating that the first two modes contain most of the energy.

Figure \ref{fig:model}(b) shows the flow rates obtained from simulations and by solving the two-dimensional reduced model [Eqs.\ (\ref{Qiai}), (\ref{dota1})-(\ref{dota2})]. The agreement between the simulations and the model is satisfactory, confirming that the reduction to two modes captures most of the dynamics. The spatial structure of these two modes can be revealed by plotting the elementary flow rates $q_{jn}$, which are cross-sectional integrals of the velocity eigenmodes $\mathbf \Phi_n$, and can be interpreted as the eigenmodes of the flow rates [Fig.~\ref{fig:model}(c)].
Mode 1 mainly corresponds to the flow oscillations in the main channels, contributing little to the oscillations in the connecting channel. In contrast, mode 2 contributes mainly to the oscillations in the connecting channel. A combination of these two modes exhibits eddies between the blades in the linking channel, corresponding to the fluid structures observed in the DNS. 

In the following section, this model is used to study the transition from steady to periodic oscillating flows.

\subsection{Onset of Hopf bifurcations and  saturation of instabilities}

We have observed numerically that the dynamical system in Eq.~(\ref{dotX})  has an equilibrium solution $\mathbf X^{\text{eq}}$. This is obtained by solving $\mathbf N(\mathbf X ,\mathbf X) + \mathbf L \mathbf (\mathbf X) + \mathbf H = \mathbf 0$ using an iterative method initiated at $\mathbf X = \mathbf 0$. The equilibrium solution corresponds to a steady flow in the elementary network. Next, we show that this flow is unstable for a range of pressures, leading to the emergence of an oscillatory mode. As the equilibrium state is perturbed, these oscillations are not periodic initially, since the amplitude grows. However, under the effect of the nonlinear terms (i.e., the terms involving the coefficients $A_{n\ell s}$ in the model), the oscillation amplitude saturates to a constant value, leading to periodic dynamics. 

The eigenvalues of the  linearized dynamics around the equilibrium point $\mathbf X^{\text{eq}}$ are complex, and thus form a complex conjugate pair $\sigma=\sigma_{\text{r}} + i \sigma_{\text{i}}$ and $\sigma^*=\sigma_{\text{r}} - i \sigma_{\text{i}}$, where the stability of the fixed point 
is determined by the real part
$\sigma_{\text{r}}$.
The results are shown in Fig.\ \ref{fig:model}(d)  for the case $P_2^{\text{in}} = 45 $ Pa, indicating that  $\sigma_{\text{r}}$ becomes positive as $P_1^{\text{in}}$ is increased past
$P_{1\text c}^{\text{in}}\approx 54 $ Pa. This marks the onset of flow rate oscillations, in agreement with Fig.\ \ref{cfd_singlechip}. 
In addition, the imaginary part $\sigma_{\text{i}}$ is observed to increase from zero and be strictly positive as soon as $P_1^{\text{in}} > P_{1\text{c}}^{\text{in}}$, which is indicative of a Hopf bifurcation.

The period and the saturated amplitude of the oscillating mode appearing in the model should correspond to those of the oscillating velocity fields in the DNS data. They can be obtained from the normal form of the system in the vicinity of the bifurcation point \cite{kuznetsov2004}. To achieve this, we set $\mathbf x = \mathbf X - \mathbf X^{\text{eq}} $, so that the model now reads:
\begin{equation}
    \dot {\mathbf x} = \mathbf N(\mathbf x,\mathbf x) + \mathbf \Lambda  \mathbf x, 
\label{dotx_Lambda}
\end{equation}
where $\mathbf \Lambda  \mathbf x = \mathbf L (\mathbf x) + \mathbf N(\mathbf X^{\text{eq}},\mathbf x) + \mathbf N(\mathbf x,\mathbf X^{\text{eq}})\nonumber$. 

Following Ref.\ \cite{kuznetsov2004}, we introduce the
eigenvector $\mathbf q$ of $\mathbf\Lambda$ corresponding to the eigenvalue $\sigma$, as well as the eigenvector $\mathbf{p}$ of $\mathbf\Lambda^T$ (adjoint eigenvector) corresponding to the eigenvalue $ \sigma^*$. These eigenvalues, examined above, describe the dynamics in the vicinity of the bifurcation point.
We also introduce the Hermitian inner product $\mathbf u\cdot \mathbf v = \sum_l u_l^* v_l$ and normalize $\mathbf p$ such that $\mathbf p \cdot \mathbf q = 1$.
One can check that $\mathbf p  \cdot  \mathbf q^* = 0$. We can then write any two-dimensional vector $\mathbf x(t)$ as $\mathbf x(t) =  z(t) \mathbf q +  z^*(t) \mathbf q^*$
with $z(t) = \mathbf p  \cdot \mathbf x $. In terms of the complex coordinate $z(t)$, the dynamical system (\ref{dotx_Lambda}) is given by
\begin{equation}\dot z = \sigma z + \frac{1}{2} g_{20}    \, z^2 + g_{11} \, z z^* + \frac{1}{2} g_{02}    \, z^{*2}, 
\end{equation}
where $g_{20}= 2 \mathbf p  \cdot 
\mathbf N(   \mathbf q  ,  \mathbf q  )$, $g_{11}=  \mathbf p \cdot 
\mathbf N(   \mathbf q  ,  \mathbf q^*  ) + \mathbf p \cdot 
\mathbf N(   \mathbf q^*  ,  \mathbf q  )$, and $g_{02}= 2 \mathbf p  \cdot 
\mathbf N(   \mathbf q^*  ,  \mathbf q^*  )$.
These coefficients can be readily computed as soon as the $A_{n \ell s}$'s are known. Finally, by applying two additional successive changes of variables to $z(t)$ \cite{kuznetsov2004}, we obtain the normal form of the model in terms of the new variable, 
\begin{equation}
\dot \zeta = \sigma \zeta + \gamma \, \zeta^2 \zeta^* + O(\zeta^4),
\label{dotzeta}
\end{equation}
where $\gamma=\gamma_{\text{r}}+i\gamma_{\text{i}}$ only depends on the $A_{n\ell s}$ and reads
$\gamma = - g_{20} \, g_{11} \! \left(\frac{1}{\sigma} + \frac{1}{2\sigma^*}\right) - \frac{1}{\sigma^*}|g_{11} |^2 - \frac{1}{2} \frac{1}{\sigma - 2\sigma^*}|g_{02}|^2$.

Equation (\ref{dotzeta}) shows that the perturbation grows with a growth rate $\sigma_{\text{r}}$ for short times, then saturates due to the nonlinear term $\gamma \zeta^2 \zeta^*$. The general solution of  this equation can be obtained by setting $\zeta(t)=\eta(t) e^{i \theta(t)}$, leading to
\begin{equation}
    \eta(t)=\left[ \left(\frac{1}{\eta(0)^2}  + \frac{\gamma_{\text{r}}}{\sigma_{\text{r}}}\right)e^{-2\sigma_{\text{r}} t} - \frac{\gamma_{\text{r}}}{\sigma_{\text{r}}}  \right]^{-1/2}. 
\end{equation}
Hence, $\zeta(t)$ is not periodic but, when $\sigma_\text{r}>0$, it converges to a periodic solution for $t \gg 1/\sigma_\text{r}$:
\begin{equation}
    \zeta (t) \approx \eta_\infty e^{i \omega_\infty t}, 
\label{zetainf}
\end{equation}
where $\eta_\infty =  \sqrt{{-\sigma_{\text{r}}}/{\gamma_{\text{r}}}} $ and 
\begin{equation}
    \omega_\infty = \frac{\gamma_{\text{r}}\sigma_{\text{i}} - \gamma_{\text{i}}\sigma_{\text{r}} }{\gamma_{\text{r}}}
\label{winf}
\end{equation}
are the amplitude and angular frequency of the saturated instability. We then obtain the 
period $T=2\pi/\omega_\infty$ of the oscillations in the microfluidic network in the nonlinear regime: 
\begin{equation}
T = \frac{2 \pi \gamma_{\text{r}} }{\gamma_{\text{r}}\sigma_{\text{i}}  - \gamma_{\text{i}} \sigma_{\text{r}} }. 
\end{equation}
We can also obtain the amplitude of the oscillations of the flow through the connecting channel for the same regime. 
Indeed, having determined $\zeta(t)$ through Eq.~(\ref{zetainf}), we reconstruct $z(t)$, then $\mathbf x(t)$, and $(a_1(t),a_2(t))^T = \mathbf X^{\text{eq}} + \mathbf x(t)$. The amplitude of the oscillations of $Q_3(t)$ is then obtained from Eq.~(\ref{Qiai}) as the amplitude of $q_{31} a_1(t) + q_{32} a_2(t) $.

The results are shown in Fig. \ref{fig:model}(e) for the angular frequency $\omega_\infty$ and Fig. \ref{fig:model}(f) for the amplitude of the flow rate in the connecting channel. We observe good agreement with the DNS in the vicinity of the bifurcation point $P_1^{\text{in}} = P_{1\text{c}}^{\text{in}}$, as expected. The curves diverge as $(P_1^{\text{in}} - P_{1\text{c}}^{\text{in}})/P_{1\text{c}}^{\text{in}}$ increases. These calculations confirm that a Hopf bifurcation takes place  and that the period and amplitude of the oscillating flow during the development of the instability (linear regime) and after the saturation (nonlinear regime and appearance of periodicity) can be reproduced using the reduced model in the vicinity of the bifurcation point.

\section{Serialization of elementary networks}
\label{compo_in_series}

Having characterized the transition to periodicity in the elementary network,
we now proceed to the case of a microfluidic network composed of $m$ elementary networks, each identical to the one considered above (Fig.~\ref{network}). These components are placed in series, where the outlets of segment $k$ are the inlets of segment $k+1$. In steady Stokes flow conditions, and for serialized two-terminal components subjected to a driving pressure $P^{\text{in}}$, one can expect that the flow structure in each component will be identical to the flow structure in a single element submitted to a driving pressure $P^{\text{in}}/m$. This ``flow-preserving serialization" is no longer guaranteed in the case of the oscillating components considered here, for two reasons. First, they have two inlets and two outlets, and the pressures 
can be different at three (or even four for $k<m$) of these terminals. Second, the system is designed to work under weak inertial conditions (i.e., moderate Reynolds numbers), rather than under Stokes flow conditions.

\begin{figure}[th!] 
    \centering
    \includegraphics[width=0.5\textwidth]{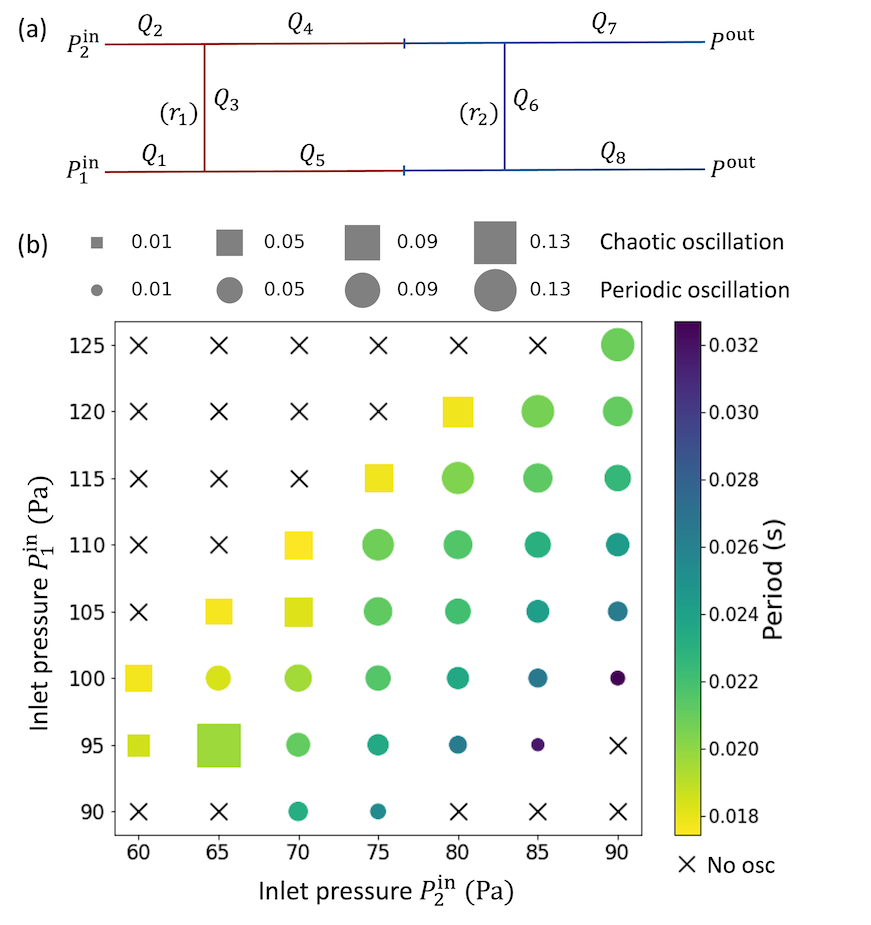}
    \caption{Analysis of serialized configuration. (a) Schematic diagram of the network serializing two segments with inlet pressures $P_1^{\text{in}}$ and $P_2^{\text{in}}$. (b) DNS results for the network in (a), where the squares and circles indicate
    chaotic and periodic oscillations. The color and size indicate the oscillation frequencies and amplitudes for the flow rate $Q_3+Q_6$ as a function of $P_1^{\text{in}}$ and $P_2^{\text{in}}$. 
    The cross sign in the upper left and lower right  indicate non-oscillatory parameter regions. }
    \label{2Hcfd_PPdomain}
\end{figure}

To examine the occurrence and development of the oscillating dynamics of the serialized systems, we now turn to DNS. 
Figure~\ref{2Hcfd_PPdomain} shows that these factors above lead to complex dynamics even in the case of $m=2$ and identical elementary networks. 
As in the case of a single elementary network, spontaneous periodic flow rate oscillations (circles) are observed for a range of pressures
in the $\left(P_1^{\text{in}},P_2^{\text{in}}\right)$ parameter space.
However, the system also exhibits chaotic flow rate oscillations (squares) for a different range of pressures, which are characterized by nonperiodic amplitudes and instantaneous frequencies.
For the chaotic cases, the plotted oscillation frequencies are dominant (peak) frequencies obtained from fast Fourier transforms, and the amplitudes are taken to be the maximum changes of flow rate.

\begin{figure*}[th]
\centering
\includegraphics[width=0.9\textwidth]{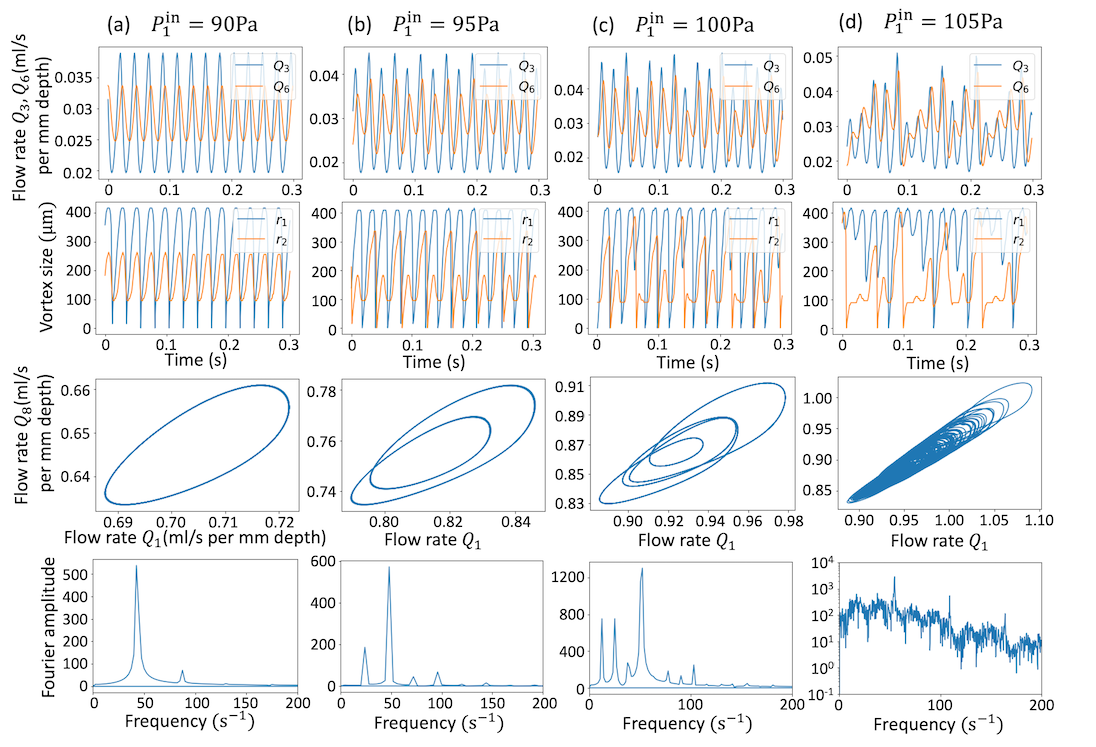}
\caption{DNS results for the two-segment network in Fig \ref{2Hcfd_PPdomain}.
(a-d) Simulations for $P_1^{\text{in}}= 90\, \text{Pa}$ (a), $95\, \text{Pa}$ (b), 100\, \text{Pa} (c), and $105\, \text{Pa}$ (d); 
all simulations are for $P_2^{\text{in}}=70\, \text{Pa}$.
Rows (top to bottom): flow-rate and vortex-size oscillations in the connecting channels as functions of time,  trajectories in the section ($Q_1$, $Q_8$) of the state space, and Fourier transform of $Q_3$. The first peaks of Fourier amplitudes in (a-c) are $f_1=42.5\, \text{s}^{-1},\, f_2=24.0\, \text{s}^{-1}$, and $f_4=13.9\, \text{s}^{-1}$, respectively.}
\label{2Hcfd}
\end{figure*}

Figure \ref{2Hcfd} shows further details of the dynamics for fixed  $P_2^{\text{in}}=70$ Pa as the inlet pressure $P_1^{\text{in}}$ is varied. 
When $P_1^{\text{in}}=90$ Pa [Fig.~\ref{2Hcfd}(a)], 
both segments exhibit simple periodic oscillations and are frequency-synchronized.  
On increasing $P_1^{\text{in}}$ to $95$, $100$, and $105$ Pa [Figs.~\ref{2Hcfd}(b-d)], 
more complex regimes are encountered through
successive period-doubling bifurcations, with chaotic dynamics occurring for  
$105\, \text{Pa}$, as anticipated in Fig~\ref{2Hcfd_PPdomain}. 
In Fig.~\ref{2Hcfd}, this is visualized in terms of the time-dependent flow rates of the connecting channels (first row) and vortex sizes (second row) as well as the
state-space trajectory (third row) and the Fourier spectra of flow rate $Q_3$ (fourth row).

Period-one dynamics is observed for $P_1^{\text{in}}=90\, \text{Pa}$ [Fig.\ \ref{2Hcfd}(a)], where the Fourier spectrum shows a peak at $f_1=42.5$ $\text{s}^{-1}$, corresponding to the period $T_1=0.024$ s, and a secondary peak associated with the harmonic $2f_1$. 
Period-two dynamics occurs for $P_1^{\text{in}}=95\, \text{Pa}$ [Fig.\ \ref{2Hcfd}(b)], where the vortex size $r_2$ oscillates alternating between two different amplitudes, $\approx\! 340\, \mu \text{m}$ and $\approx\! 190\, \mu \text{m}$,  while the vortex size $r_1$ oscillates with a fixed amplitude of $\approx\! 400\, \mu \text{m}$ [see Fig.~\ref{2Hcfd_PPdomain}(a) for notation]. 
This indicates that the oscillations in the second segment are emergent and no longer dominantly driven by the first segment, even though the connecting channel oscillations exhibit two amplitudes in both segments. Here, the spectrum shows a first peak at $f_2=24.0$ s$^{-1}$, with a period $T_2=0.042$ s, which is nearly twice $T_1$, and a main peak at $2f_2$.
Period-four dynamics is observed for $P_1^{\text{in}}=100\, \text{Pa}$ [Fig.\ \ref{2Hcfd}(c)], where lower frequencies emerge in the spectrum, with the first major peak appearing at $f_4=13.9$ s$^{-1}$.  
Finally, we observe that when $P_1^{\text{in}}=105\, \text{Pa}$ [Fig.\ \ref{2Hcfd}(d)] the dynamics have very complex features and a wide spectrum, both consistent with the presence of chaos.

\begin{figure}[t]
\centering
\includegraphics[width=0.5\textwidth]{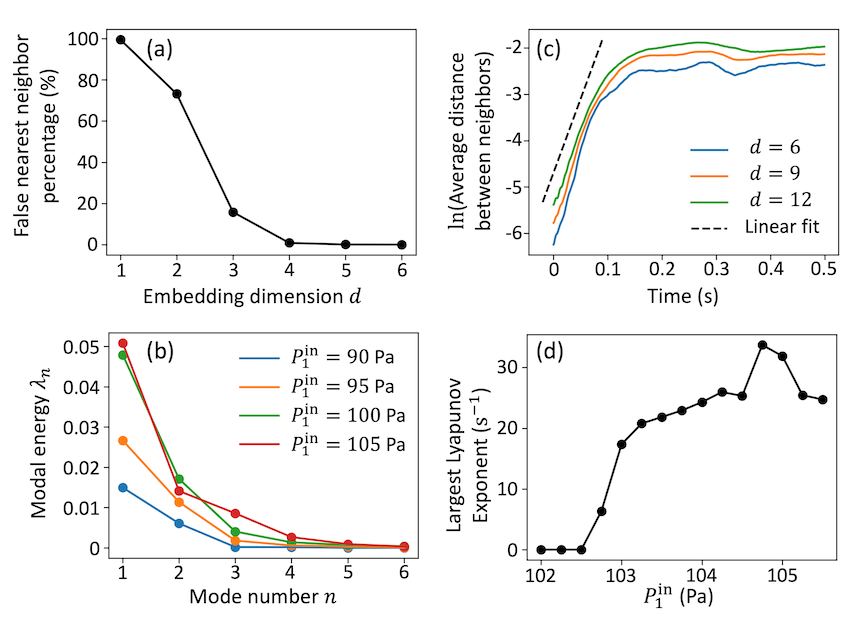}
\caption{Dominant degrees of freedom and Lyapunov exponents for the two-segment network in Fig.~\ref{2Hcfd_PPdomain} with $P_2^{\text{in}}=70\, \text{Pa}$.
(a) False nearest neighbor percentage vs.\  embedding dimension for $P_1^{\text{in}}=105\, \text{Pa}$.
(b) Modal energies (non-dimensional) from the reduced model for various $P_1^{\text{in}}$. (c) Average distance between pairs of neighbors for $P_1^{\text{in}}=105\, \text{Pa}$ with different embedding dimensions $d$. The dashed line is a linear least-squares fit of the $d=9$ curve, whose slope is an estimation of the largest Lyapunov exponent. 
(d) Largest Lyapunov exponent of the system estimated for a range of values of $P_1^{\text{in}}$.
The calculations are based on the reduced model (b) and time series data from DNS (other panels).}
\label{lyapunov}
\end{figure}

We confirm the occurrence of chaos by calculating the largest Lyapunov exponent from {DNS} data using the local divergence rate in the phase space \cite{kantz1994robust}, which is more accurate than using the reduced model but requires determining the relevant embedding dimension $d$.
To obtain the lowest embedding dimension needed to unfold the attractor, we calculate the false nearest neighbor percentage vs.\ embedding dimension \cite{kennel1992determining} using the time series of $Q_1$, $Q_2$, and $Q_8$ from DNS data. 
As shown in Fig.~~\ref{lyapunov}(a), this percentage decreases rapidly with $d$, is almost zero for $d=4$, and vanishes when $d>8$ for the conditions in Fig.\ \ref{2Hcfd}(d).
Using the POD decomposition, we also extract the energy of the dominant modes [Fig.~\ref{lyapunov}(b)]. 
For the periodic case $P_1^{\text{in}} = 90$ Pa, 
we observe that two modes are sufficient to describe the dynamics.
However, as the flow complexity increases for larger $P_1^{\text{in}}$,  additional modes are required. For the chaotic case  $P_1^{\text{in}}=105$ Pa, four modes are needed, which is in agreement with the estimate of the lowest embedding dimension.

We then reconstruct the flow-rate data using a few different embedding dimensions and choose $d=9$ to calculate the Lyapunov exponent. 
In the reconstructed state space, we consider the logarithm of the average distance between pairs of neighboring points over time for a large number of pairs [approximately $1000$ in the numerics shown in Fig.~\ref{lyapunov}(c)]. 
This function increases almost linearly before reaching the overall length scale of the reconstructed attractor, and a least-squares  fit over the linear portion estimates the largest Lyapunov exponent to be approximately $31.9$ s$^{-1}$.
This positive value confirms that the system is chaotic for $P_1^{\text{in}}=105\, \text{Pa}$ and $P_2^{\text{in}}=70\, \text{Pa}$.

Similar results are obtained for other pressures $P_1^{\text{in}}$ larger than approximately $103\, \text{Pa}$. The results are shown in Fig.\ \ref{lyapunov}(d), where we use an embedding dimension of $d=12$, which we verified to be 
adequate for the entire range of pressures considered.
These results further elucidate the emergence of a chaotic attractor as a result of period doubling. Crucially, as anticipated above and shown in Fig.\ \ref{2Hcfd}(d), in this regime serialization does not preserve the spatiotemporal structure of the flow, indicated by the markedly different oscillation patterns in the two network segments. Moreover, synchronization between the oscillations in the elementary networks becomes less pronounced, 
reflecting the increasingly irregular
amplitude behavior within each network.

\section{Concluding remarks}
\label{conclusion}
\color{black} 

Our numerical simulations and modal analysis characterize the emergence of complex flow dynamics in two-dimensional microfluidic networks composed of simple H-shaped structures with blade-like obstacles in their connecting channels. By only controlling the constant driving pressures at the two inlets, we observe the spontaneous emergence of persistent flow rate oscillations associated with the time-dependent behavior of vortices near the obstacles. These oscillations are not due to any compliance of the channel or volume change, as the walls are rigid and the fluid is incompressible. They are instead due to the finite inertia of the fluid, which induces nonlinearities in the equations of motion, and its interplay with the geometry of the network.
For a range of moderate Reynolds numbers, the system yields a ``microfluidic clock'' with highly regular oscillations solely due to the hydrodynamics of the flow, foregoing the need for external periodic forcing.

This phenomenon is markedly distinct from well-established instabilities in systems involving flow past a cylindrical obstacle,
which also lead to time-varying vortices but
require higher Reynolds numbers in confined systems and do not lead to significant flow rate oscillations \cite{ooi2020some}.

Crucially, the emergence of spontaneous oscillations is observed for a wide range of driving pressures, leading to diverse oscillation frequencies and amplitudes. By allowing different inlet pressures (and thus effectively treating the system as a three-terminal device), we were able to demonstrate that such oscillations are achieved even in the absence of any nonlinearity outside the connecting channel.
This generalizes previous findings from systems with stronger nonlinearities
\cite{case2020spontaneous}, where identical inlet pressures were considered (rendering the system a two-terminal device).
The modal analysis shows that the dynamics of our system can be described by a small number of degrees of freedom, giving rise to a dimension-reduced dynamical model that is amenable to analytical treatment.
Nonlinear stability analysis of this model allows the derivation of the oscillation frequencies and amplitudes in the periodic regime. 

Insight into the persistence of spontaneous oscillations in larger networks was derived by considering two elementary networks connected in series. Because the system has three terminals and the inertial forces are non-negligible, the flow structure is generally not preserved across the elementary networks in the serialized system. Notwithstanding, for certain pressure ranges, spontaneous flow rate oscillations persist and remain frequency-synchronized across the two networks. 
However, by varying the inlet pressures, the system can undergo a series of period-doubling bifurcations. This results in substantially more complex dynamics and a larger number of effective degrees of freedom, eventually leading to chaotic flow rates. Physically, the emergence of chaos is related to the chaotic dynamics of vortices in the connecting channels. 
Note that, even though Lagrangian chaos in Stokes flow is a very common phenomenon that led to the concept of micromixing \cite{ottino2004designing}, the chaotic dynamics observed here are of a different nature. In this case, the irregular behavior is reflected in the flow rates and can be interpreted as a nonturbulent instance of Eulerian chaos. 

An important direction for future research is to study these phenomena experimentally. To reproduce our predictions from two-dimensional simulations, 
experiments should be designed with a channel depth significantly larger than the channel width. 
Experiments with shallow channels would also be  informative, as they could reveal important depth-dependent flow structures. The latter points to the significance of analyzing three-dimensional models in future research. 

The findings presented here demonstrate the rich dynamics achievable in unmodulated systems with rigid, non-compliant channels and advance the fundamental understanding of inertial effects in microfluidic networks.
This understanding can in turn open a door to potential applications in precision fluid control, microfluidic timing devices, and chaos-based  technologies.

\section*{Acknowledgement}
The authors thank Daniel J.\ Case for insightful discussions and acknowledge the use of computer resources from  Northwestern University's High Performance Computing Cluster. 
This work was supported by ARO Grant No.\ W911NF-20-1-0173.

\end{document}